\documentclass{article}
\usepackage{amsmath}
\usepackage{amsthm}
\usepackage{amssymb,epsfig}
\usepackage[cp 1250]{inputenc}
\usepackage{a4}
\usepackage{graphics}

\newtheorem{lemma}{Lemma}[section]

\newtheorem{thm}[lemma]{Theorem}
\theoremstyle{definition}

\newtheorem{example}[lemma]{Example}

\newtheorem{proposition}[lemma]{Proposition}
\newtheorem{corollary}[lemma]{Corollary}
\newtheorem{observation}[lemma]{Observation}

\def\Z{\mathbb Z}
\def\N{\mathbb N}
\def\A{\mathcal A}

\def\pf{\begin{proof}}
\def\pfk{\end{proof}}

\begin{document}
\title{Combinatorial and Arithmetical Properties of Infinite Words Associated with Non-simple Quadratic Parry Numbers}

\author{Lubom\'ira Balkov\'a, \and
Edita Pelantov\'a\footnote{pelantova@km1.fjfi.cvut.cz}, \and Ond\v
rej Turek }

\maketitle \thispagestyle{empty}

\begin{center}
Doppler Institute for Mathematical Physics and Applied Mathematics and\\
Department of Mathematics, FNSPE, Czech Technical University, \\
Trojanova 13, 120 00 Praha 2, Czech Republic
\end{center}

\begin{abstract}
We study   arithmetical and combinatorial properties  of
$\beta$-integers
 for $\beta$
being the root of the equation $x^2=mx-n, m,n \in \mathbb N, m
\geq n+2\geq 3$.  We determine with the accuracy of $\pm 1$ the
maximal number of $\beta$-fractional positions, which may arise as
a~result of addition of two $\beta$-integers. For the infinite
word  $u_\beta$ coding distances  between consecutive
$\beta$-integers, we determine precisely  also the balance. The
word $u_\beta$ is the fixed point of the morphism  $A \to
A^{m-1}B$ and $B\to A^{m-n-1}B$.
 In
the case  $n=1$ the corresponding infinite word $u_\beta$ is
sturmian and therefore $1$-balanced. On the simplest non-sturmian
example with $n\geq 2$, we illustrate how closely the balance and
arithmetical  properties of $\beta$-integers are related.
\end{abstract}

\section{Introduction}
In this paper, we focus on study of arithmetical and combinatorial
properties of $\beta$-integers for $\beta$ being a~quadratic
algebraic integer with positive norm. The notion of $\beta$-integer
is related to greedy algorithm searching for the expansion of a~real
number $x$ in base $\beta >1$; this algorithm has been introduced
in~\cite{Renyi} by R\'enyi. The real number $x$ is called
$\beta$-integer if its $\beta$-expansion has the form $\pm
\sum_{k=0}^{n} x_k \beta^{k}$, i.e. if all of its coefficients at
powers $\beta^{-k}$ vanish for $k>0$. The set of $\beta$-integers
(denoted by $\mathbb Z_{\beta}$) equals in case of $\beta \in
\mathbb N$ to the set of integers $\mathbb Z$. If $\beta$ is not an
integer, the set $\mathbb Z_{\beta}$ has much more interesting
properties:
\begin{enumerate}
\item
$\mathbb Z_{\beta}$ is not invariant under translation.
\item
$\mathbb Z_{\beta}$ has no accumulation points.
\item
$\mathbb Z_{\beta}$ is relatively dense (= distances between
successive elements of $\mathbb Z_{\beta}$ are bounded).
\item
$\mathbb Z_{\beta}$ is self-similar, i.e. $\beta \mathbb
Z_{\beta}\subset \mathbb Z_{\beta}$.
\end{enumerate}
After the discovery of quasicrystals in $1982$ \cite{Shechtman}, it
has turned out that the set $\mathbb Z_{\tau}$, where
$\tau=\frac{1+\sqrt{5}}{2}$ is the golden mean, serves as a~model
describing coordinates of atoms in these materials with long-range
orientational order and sharp diffraction images of
non-crystallographic $5$-fold symmetry. Later on, quasicrystals with
other non-crystallographic symmetries have been found. In order to
serve as a~convenient model for quasicrystals, the set $\mathbb
Z_{\beta}$ must satisfy together with conditions $1.-4.$ also
another natural property, the so-called finite local complexity. In
one-dimensional case, it means that there exist only a~finite number
of types of distances between successive elements of $\mathbb
Z_{\beta}$. From results \cite{Thurston, Parry}, it follows that
$\mathbb Z_{\beta}$ has this property if and only if the R\'enyi
expansion of unity in base $\beta$ is eventually periodic. Such
numbers $\beta$ are called Parry numbers. It can be easily shown
that every Parry number $\beta$ is an algebraic integer, i.e. it is
a~root of a~monic polynomial having integer coefficients. The task
to describe which algebraic integers are Parry numbers has not been
solved yet. It is known that each Pisot number is as well Parry. Let
us remind that an algebraic integer $\beta$ is a~Pisot number if all
of its conjugates have modulus less than $1$. In case of $\beta$
being a~Pisot number, $\beta$-integers form the Meyer set, i.e. it
holds
$$\mathbb
Z_{\beta}-\mathbb Z_{\beta} \subset \mathbb Z_{\beta}+F$$ for
a~finite set $F \subset \mathbb R$. Thus, the notion of Meyer set
generalizes the notion of lattice, which is crucial for
description of crystals. As we have already mentioned, in case of
$\beta$ being a~Parry number, the set $\mathbb Z_{\beta}$ disposes
with a~finite number of distances between neighbors. If we
associate to different gaps different letters, it is possible to
encode the set $\mathbb Z_{\beta}$ as an infinite word $u_{\beta}$
over a~finite alphabet. Combinatorial properties of words
$u_{\beta}$ have been studied in several papers: \cite{FrMaPe1,
FrMaPe2} is devoted to description of factor complexity of
$u_{\beta}$, palindromes of $u_{\beta}$ are described in
papers~\cite{AmFrMaPe1, Ba}. So far the least studied problem is
the balance of $u_{\beta}$, i.e. the maximal difference in numbers
of different letters in factors of the same length. Balance is
clearly known for $\mathbb Z_{\beta}$ which corresponds to
sturmian words, i.e. for $\beta$ being a~quadratic unit.
In~\cite{Tu}, the balance property for $u_\beta$, where $\beta$ is
the larger root of the quadratic polynomial $x^2-mx-n,$ $ m,n \in
\mathbb N,$ $ m\geq n\geq 1,$ has been studied. For other types of
irrationalities, the balance property has not been described yet.

The sets of ordinary integers and $\beta$-integers are very
different also from the arithmetical point of view. $\mathbb
Z_{\beta}$ is not closed under addition and multiplication for any
$\beta \not \in \mathbb N$. Sum of two $\beta$-integers may even
not have a~finite $\beta$-expansion. So far unsolved and likely
very difficult is the question of characterization of those
$\beta$ for which this pathological situation does not appear.
Mathematically expressed it means to describe $\beta$ for which
the set $Fin(\beta)$, i.e. the set of numbers with finite
$\beta$-expansion, is a~subring of $\mathbb R$. Frougny and
Solomyak have shown in~\cite{FrSo} that the necessary condition
for this so-called finiteness property is that $\beta$ is a~Pisot
number. Some sufficient conditions can be found in~\cite{Akiyama,
FrSo, Hollander}. If sum or product of two $\beta$-integers has
a~finite $\beta$-expansion, there arises a~question how long is
the $\beta$-fractional part of the sum or product. This problem
has been investigated in~\cite{AmFrMaPe2, Bernat, BuFrGaKr,
GuMaPe, Messaoudi}.

Here, the main attention is devoted to investigation of
arithmetics of $\beta$-integers for a~non-simple quadratic Parry
number $\beta$, i.e. for $\beta$ being the root of the equation
$x^2=mx-n, m,n \in \mathbb N, m \geq n+2\geq 3$. We determine with
the accuracy of $\pm 1$ the maximal number of $\beta$-fractional
positions $L_{\oplus}(\beta)$, which may arise as a~result of
addition of two $\beta$-integers. So we improve considerably the
estimate from the paper~\cite{GuMaPe}. We determine accurately
also the balance of $u_\beta$. On this easiest non-sturmian
example, we illustrate how closely the arithmetical and
combinatorial properties of $\mathbb Z_{\beta}$ are related.
Particulary, we show the relation between $L_{\oplus}(\beta)$ and
the balance property. Our method might be applied also for
determination of the balance property for words coding
$\beta$-integers with irrationalities of a~higher degree.

\section{Preliminaries}\label{Preliminaries}
An {\em alphabet} $\cal A$ is a~finite set of symbols called {\em
letters}. A~concatenation of letters is a~{\em word}. The set $\cal
A^{*}$ of all finite words (including the empty word $\varepsilon$)
provided with the operation of concatenation is a~free monoid. The
length of a~word $w=w_0w_1w_2\cdots w_{n-1}$ is denoted by $|w|=n$.
We will deal also with infinite words $u=u_0u_1u_2\cdots$. A~finite
word $w$ is called a~{\em factor} of the word $u$ (finite or
infinite) if there exist a~finite word $w^{(1)}$ and a~word
$w^{(2)}$ (finite or infinite) such that $u=w^{(1)}ww^{(2)}$. The
word $w$ is a~{\em prefix} of $u$ if $w^{(1)}=\varepsilon$.
Analogically, $w$ is a~{\em suffix} of $u$ if $w^{(2)}=\varepsilon$.
A~concatenation of $k$ words $w$ will be denoted by $w^k$,
a~concatenation of infinitely many finite words $w$ by $w^{\omega}$.
An infinite word $u$ is said to be {\em eventually periodic} if
there exist words $v,w$ such that $u=v w^{\omega}$. A~word which is
not eventually periodic is called {\em aperiodic}. We will denote by
$\mathcal{L}(u)$ (language on $u$) the set of all factors of the
word $u$. $\mathcal{L}_n(u)$ denotes the set of all factors of
length $n$ of the word $u$, clearly
$${\mathcal L}(u)=\bigcup_{n \in \mathbb
N}{\mathcal L}_n(u).$$

The measure of variability of local configurations in $u$ is
expressed by the factor complexity function ${\cal C}_u:\mathbb N
\rightarrow \mathbb N$, which associates with $n \in \mathbb N$ the
number ${\cal C}_u(n):=\#{\mathcal L}_n(u)$. Obviously, a~word $u$
is eventually periodic if and only if ${\cal C}_u(n)$ is bounded by
a~constant. On the other hand, one can show that a~word $u$ is
aperiodic if and only if ${\cal C}_u(n) \geq n+1$ for all $n \in
\mathbb N$. Infinite aperiodic words with the minimal complexity
${\cal C}_u(n)=n+1$ for all $n \in \mathbb N$ are called sturmian
words. These words are studied intensively, several different
definitions of sturmian words can be found in \cite{Berstel}.

Another way how to measure the degree of variability in the
infinite word $u$ is the balance property. Let us denote the
number of letters $a \in \cal A$ in the word $w$ by $|w|_{a}$. We
say that an infinite word $u$ is \mbox{$c$-balanced}, if for every
$a \in \cal A$ and for every pair of factors $w$, $\hat{w}$ of
$u$, with the same length $|w|=|\hat{w}|$, we have
$\left||w|_{a}-|\hat{w}|_{a}\right|\leq c$. Note that in the case
of binary alphabet $\A=\{A,B\}$, this condition may be written in
the simpler way as $\left||w|_A-|\hat{w}|_A\right|\leq c$.
Sturmian words are characterized by the property that they are
$1$-balanced (or simply balanced)\cite{Morse}. To determine the
minimal constant $c$ for which the infinite word is $c$-balanced
is a~difficult task. Adamczewski gives an upper bound on $c$ for
a~certain class of infinite words. To describe his result, we must
introduce the notion of morphism. A~mapping $\varphi$ on the free
monoid $\cal A^{*}$ is called a~morphism if
$\varphi(vw)=\varphi(v)\varphi(w)$ for all $v,w \in \cal A^{*}$.
Obviously, for determining the morphism it suffices to give
$\varphi(a)$ for all $a \in \cal A$. The action of the morphism
can be naturally extended on right-sided infinite words by the
prescription
$$\varphi(u_0u_1u_2\cdots):=\varphi(u_0)\varphi(u_1)\varphi(u_2)\cdots$$
A~non-erasing morphism $\varphi$, for which there exists a~letter $a
\in \cal A$ such that $\varphi(a)=aw$ for some non-empty word $w \in
\cal A^{*}$, is called a~substitution. An infinite word $u$ such
that $\varphi(u)=u$ is called a~fixed point of the substitution
$\varphi$. Obviously, every substitution has at least one fixed
point, namely
$$\lim_{n \to \infty}\varphi^{n}(a).$$

To any substitution $\varphi$ on the $k$-letter alphabet ${\cal
A}=\{a_1, a_2,...,a_k \}$, one can associate the so-called {\it
incident matrix} $M$ of size $k \times k$ defined by
$$M_{ij}:=|\varphi(a_i)|_{a_j}.$$
The result of Adamczewski concerns infinite words $u$ being fixed
points of primitive substitutions. Recall that a~substitution
$\varphi$ is primitive if there exists a~power $k$ of $\varphi$ such
that each pair of letters $a,b \in {\cal A}$ satisfies
$|\varphi^{k}(a)|_b \geq 1$. In accordance with the Perron-Frobenius
theorem, the incident matrix of a~primitive substitution has one
real eigenvalue greater than one, which is moreover greater than the
modulus of all the other eigenvalues. This eigenvalue, say
$\Lambda$, is called the Perron eigenvalue of the substitution. In
\cite{Adamczewski} it has been proved that if $u$ is the fixed point
of a~primitive substitution with the incidence matrix $M$, then $u$
is $c$-balanced for some constant $c$ if and only if $|\lambda| < 1$
for all eigenvalues $\lambda$ of $M$, $\lambda \not =\Lambda$.

\subsection{Beta-expansions and beta-integers} \label{beta}
Let $\beta >1$ be a~real number and let $x$ be a~positive real
number. Any convergent series of the form:
$$x=\sum_{i=-\infty}^{k} x_i\beta^{i},$$
where $x_i \in \mathbb N$, is called a~{\em $\beta$-representation}
of $x$. As well as it is usual for the decimal system, we will
denote the $\beta$-representation of $x$ by
$$x_k x_{k-1}\cdots x_0 \bullet x_{-1}\cdots\qquad   \ \hbox{if }\  k\geq 0,$$
and
$$0 \bullet \!\!\!\!\!\!\underbrace{0\cdots0}_{(-1-k)-times} \!\!\!\!\!\!x_{k}x_{k-1}\cdots\qquad \hbox{otherwise.}$$
If a~$\beta$-representation ends with infinitely many zeros, it is
said to be finite and the ending zeros are omitted. If $\beta \not
\in \mathbb N$, for a~given $x$ there can exist more
$\beta$-representations.

Any positive number $x$ has at least one representation. This
representation can be obtained by the following  {\it greedy
algorithm:}

\begin{enumerate}
\item Find  $k \in \mathbb Z$ such that $\beta^{k} \leq x <
\beta^{k+1}$ and put $x_k:=\lfloor \frac{x}{\beta^{k}}\rfloor$ and
$r_k:=\{ \frac{x}{\beta^{k}}\}$, where $\lfloor x \rfloor$ denotes
the lower integer part and $\{x\}=x-\lfloor x \rfloor$ denotes the
fractional part of $x$.

\item For $i<k$, put $x_i:=\lfloor \beta r_{i+1}\rfloor$ and
$r_i:=\{\beta r_{i+1}\}$.
\end{enumerate}

 The representation obtained by the
greedy algorithm is called {\em $\beta$-expansion} of $x$ and the
coefficients of a~$\beta$-expansion clearly satisfy: $x_{k} \in
\{1,\ldots,\lceil \beta \rceil -1\}$ and $x_i \in
\{0,\ldots,\lceil \beta \rceil-1\}$ for all $i<k$, where $\lceil x
\rceil$ denotes the upper integer part of $x$. We will use for
{\em $\beta$-expansion} of $x$ the notation ${\langle x \rangle}
_{\beta}.$ If $x=\sum_{i=-\infty}^{k} x_i\beta^{i}$ is the
$\beta$-expansion of a~nonnegative number $x$, then
$\sum_{i=-\infty}^{-1} x_i\beta^{i}$ is called the
$\beta$-fractional part of $x$. Let us introduce some important
notions connected with $\beta $-expansions:
\begin{itemize}
\item The set of nonnegative numbers with vanishing $\beta$-fractional part
are called nonnegative $\beta$-integers, formally
$${\mathbb
Z}_{\beta}^{+}:=\{x \geq 0 \bigm | {\langle x \rangle} _{\beta}=x_k
x_{k-1}\cdots x_0 \bullet \}.$$
\item The set of $\beta$-integers is then defined by
$${\mathbb Z}_{\beta}:=\bigl( -{\mathbb Z}_{\beta}^{+}\bigr) \cup
{\mathbb Z}_{\beta}^{+}. $$
\item All the real numbers with a~finite
$\beta$-expansion of $|x|$ form the set $Fin(\beta)$, formally
$$Fin(\beta):=\bigcup_{n\in \mathbb N}\frac{1}{{\beta}^n} \mathbb Z_{\beta}.$$
For any $x \in Fin(\beta)$, we denote by $fp_{\beta}(x)$ the length
of its fractional part, i.e.
$$fp_{\beta}(x)=\min\{l \in \mathbb N \bigm | \beta^l x \in \mathbb Z_{\beta}\}.$$
\end{itemize}

The sets ${\mathbb Z}_{\beta}$ and $Fin(\beta)$ are generally not
closed under addition and multiplication. The following notion is
important for studying of lengths of the fractional parts which may
appear as a result of addition and multiplication.
\begin{itemize}
\item $L_{\oplus}(\beta):=\min \{L \in \mathbb N \bigm | x,y \in {\mathbb Z}_{\beta},
x+y \in Fin(\beta) \Longrightarrow
fp_{\beta}(x+y) \leq L \}$.
\item
$L_{\otimes}(\beta):=\min \{L \in \mathbb N \bigm | x,y \in {\mathbb
Z}_{\beta}, xy \in Fin(\beta) \Longrightarrow fp_{\beta}(xy) \leq L
\}$.
\end{itemize}
If such $L \in \mathbb N$ does not exist, we set
$L_{\oplus}(\beta):=\infty$ or $L_{\otimes}(\beta):=\infty$.

The R\' enyi expansion of unity simplifies description of elements
of ${\mathbb Z}_{\beta}$ and $Fin(\beta)$. For its definition, we
introduce the transformation $T_{\beta}(x):=\{\beta x\}$ for $x \in
\left[ 0,1 \right]$. The {\em R\'enyi expansion of unity} in base
$\beta$ is defined as
$$d_{\beta}(1)=t_1t_2t_3\cdots, \quad \mbox{where }\quad t_i:=\lfloor
\beta T_{\beta}^{i-1}(1)\rfloor.$$

Every number $\beta>1$ is characterized by its R\'enyi expansion of
unity. Note that $t_1=\lfloor \beta\rfloor \geq1$. Not every
sequence of nonnegative integers is equal to $d_{\beta}(1)$ for some
$\beta$. Parry studied this problem in his paper~\cite{Parry}:
A~sequence $(t_i)_{i \geq 1}$, $t_i \in \mathbb N$, is the R\'enyi
expansion of unity for some number $\beta$ if and only if the
sequence satisfies
$$t_jt_{j+1}t_{j+2}\cdots\prec t_1t_2t_3\cdots\quad \mbox{for every $j >1$,}$$
where $\prec$ denotes strictly lexicographically smaller.

The R\'enyi expansion of unity enables us to decide whether a~given
$\beta$-representation of $x$ is the $\beta$-expansion or not. For
this purpose, we define the infinite R\' enyi expansion of unity
\begin{equation}\label{infinite_exp}
d^{*}_{\beta}(1)= \left\{\begin{array}{ll} d_{\beta}(1) & \hbox{if}
\quad d_{\beta}(1) \quad \hbox{is infinite}\\
(t_1 t_2\cdots t_{m-1}(t_m -1))^{\omega}
 & \hbox{if}
\quad d_{\beta}(1)=t_1\dots t_m \quad \hbox{with} \quad t_m \not = 0
\end{array}\right.
\end{equation}
Parry has proved also the following proposition.
\begin{proposition} \label{Parry_expansions}
Let $d^{*}_{\beta}(1)$ be an infinite R\'enyi expansion of unity.
Let $\sum_{i=-\infty}^{k} x_i\beta^{i}$ be a~$\beta$-representation
of a~positive number $x$. Then $\sum_{i=-\infty}^{k} x_i\beta^{i}$
is a~$\beta$-expansion of $x$ if and only if $x_ix_{i-1}\cdots \prec
d^{*}_{\beta}(1)$ for all $i \leq k$.
\end{proposition}

\subsection{Infinite words associated with $\beta$-integers}
If $\beta$ is an integer, then clearly $\mathbb Z_{\beta}=\mathbb Z$
and the distance between neighboring elements of $\mathbb Z_{\beta}$
for a~fixed $\beta$ is always 1. The situation changes dramatically
if $\beta \not \in \mathbb N$. In this case, the number of different
distances between neighboring elements of $\mathbb Z_{\beta}$ is at
least 2. In \cite{Thurston}, it is shown that the distances
occurring between neighbors of $\mathbb Z_{\beta}$ form the set
$\{\Delta_{k}\bigm | k \in \mathbb N \}$, where
\begin{equation}\label{distance}
\Delta_{k}:=\sum_{i=1}^{\infty}\frac{t_{i+k}}{{\beta}^{i}} \ \ \
\hbox{for} \ k \in \mathbb N\,.
\end{equation}
It is evident that the set $\{\Delta_{k}\bigm | k \in \mathbb N \}$
is finite if and only if $d_{\beta}(1)$ is eventually periodic.

When $d_{\beta}(1)$ is eventually periodic, we will call $\beta$
a~{\em Parry number}. When $d_{\beta}(1)$ is finite, it is said to
be a~{\em simple Parry number}. Every Pisot number, i.e. a~real
algebraic integer greater than 1, all of whose conjugates are of
modulus strictly less than 1, is a Parry number~\cite{Bertrand}.

From now on, we will restrict our considerations to the quadratic
Parry numbers. The R\'enyi expansion of unity for a~simple
quadratic Pisot number $\beta$ is equal to $d_{\beta}(1)=pq$,
where $p \geq q$. Hence, $\beta$ is exactly the positive root of
the polynomial $x^2-px-q$. Whereas the R\'enyi expansion of unity
for a~non-simple quadratic Pisot number $\beta$ is equal to
$d_{\beta}(1)=pq^{\omega}$, where $p>q \geq 1$. Consequently,
$\beta$ is the greater root of the polynomial $x^2-(p+1)x+p-q$.
Drawn on the real line, there are only two distances between
neighboring points of $\Z_\beta$.  The longer distance is always
$\Delta_0=1$, the smaller one is $\Delta_1$. Conversely, if there
are exactly two types of distances between neighboring points of
$\Z_\beta$ for $\beta>1$, then $\beta$ is a quadratic Pisot
number.

If we assign letters $A$, $B$ to the two types of distances
$\Delta_0$ and $\Delta_1$, respectively, and write down the order of
distances in $\Z^{+}_{\beta}$ on the real line, we naturally obtain
an infinite word; we will denote this word by $u_{\beta}$. Since
$\beta \Z^{+}_{\beta} \subset \Z^{+}_{\beta}$, it can be shown
easily that the word $u_{\beta}$ is a fixed point of a certain
substitution $\varphi$ (see~\cite{Fabre}); in particular,
for the simple quadratic Pisot number $\beta$, the generating
substitution is
\begin{equation} \label{subst1}
\varphi(A)=A^pB,\quad \varphi(B)=A^q,
\end{equation}
for the non-simple quadratic Pisot number $\beta$, the generating
substitution is
\begin{equation} \label{subst2}
\varphi(A)=A^pB,\quad \varphi(B)=A^qB.
\end{equation}

Let us remark that the matrices of these substitutions are $
 \left ( \begin{array}{cc}
 p & 1 \\
 q & 0
 \end{array}\right )$
 and
 $\left ( \begin{array}{cc}
 p  &  1 \\
 q  &  1
  \end{array} \right )$, respectively, i.e. both  substitutions
  are primitive. Therefore it follows from result \cite{Adamczewski} that there exists $c$
such that
 $u_{\beta}$ is $c$-balanced.

  In the case of $\beta$ being the root of
 $x^2-px-q$, i.e. $\beta$  quadratic simple Parry number,  the smallest possible constant $c$ was found: In \cite{Tu}
  it is  shown   that the infinite word
 generated by substitution~(\ref{subst1}) is $\left(1+\lfloor(p-1)/(p+1-q)\rfloor\right)$-balanced.
Also the
 values of $L_{\oplus}(\beta)$ have been quite precisely estimated in
 \cite{GuMaPe}:
 $$L_{\oplus}(\beta)=2p\,, \quad \mbox{if} \quad q=p; $$
 $$2 \Bigl\lfloor \frac{p+1}{p-q+1} \Bigr\rfloor \leq L_{\oplus}(\beta)
 \leq 2 \Bigl\lceil \frac{p}{p-q+1}\Bigr\rceil\,, \quad \mbox{if} \quad q<p.$$

 In this paper, we consider therefore the arithmetical properties of
 $\mathbb Z_{\beta}$ and associated infinite words $u_{\beta}$ for
 $\beta$ being the larger root of the equation $x^2-(p+1)x+p-q$.

\section{Beta-arithmetics for non-simple quadratic Parry number}\label{arithbeta}
The aim of this section is to improve the upper bound on the number
$L_{\oplus}(\beta)$ for $\beta$ having the R\'enyi expansion of
unity equal to $d_{\beta}(1)=pq^{\omega}$ for $q \leq p-1$. In the
case of $q=p-1$, $\beta$ is the larger root of the equation
$x^2-(p+1)x+1=0$, thus $\beta$ is a~quadratic unit. For
quadratic units in \cite{BuFrGaKr}, it is shown that
$L_{\oplus}(\beta)=L_{\otimes}(\beta)=1$. Let us focus on the case
of $q<p-1$. In \cite{GuMaPe}, one can find the following
estimates:
$$L_{\oplus}(\beta) \leq 3(p+1)\ln (p+1) \quad  \mbox{and} \quad  L_{\otimes}(\beta) \leq 4(p+1)\ln (p+1).$$
Here, the estimate on $L_{\oplus}(\beta)$ will be improved.
In~\cite{FrSo} and in~\cite{AmFrMaPe2}, it is shown that if
$d_{\beta}(1)=t_1t_2\cdots t_m(t_{m+1})^{\omega}$ and $ t_1\geq
t_2\geq\cdots\geq t_m >t_{m+1}$, then $Fin(\beta)$ is closed under
addition of positive elements. This fact implies that if a~number
$x$ has a~certain finite $\beta$-representation, then $x$ has as
well finite $\beta$-expansion. It follows from the definition of
greedy algorithm that if $x_kx_{k-1}\cdots x_0\bullet
x_{-1}x_{-2}\cdots$ is the $\beta$-expansion of $x>0$ and
$\tilde{x}_k\tilde{x}_{k-1}\cdots\tilde{x}_0\bullet
\tilde{x}_{-1}\tilde{x}_{-2}\cdots$ is a~$\beta$-representation of
$x$, then
$$\tilde{x}_k\tilde{x}_{k-1}\cdots\tilde{x}_0
\tilde{x}_{-1}\tilde{x}_{-2}\cdots\preceq x_kx_{k-1}\cdots x_0
x_{-1}x_{-2}\cdots$$ Thus, the $\beta$-expansion of $x$ is the
lexicographically greatest $\beta$-representation of $x$.

Let us limit our considerations to the special case of
$d_{\beta}(1)=pq^{\omega}$. The shortest and lexicographically
smallest words that do not fulfill the Parry condition are the
words
$$(p+1) \ \mbox{and} \ pq^s(q+1), \ \mbox{where} \ s \geq 0.$$

Using the equation ${\beta}^2=(p+1)\beta-(p-q)$, one can easily
obtain:
\begin{equation}\label{1}
(p+1)\bullet=10\bullet (p-q)
\end{equation}
\begin{equation}\label{2}
pq^{s}(q+1)\bullet=10^{s+2}\bullet (p-q)
\end{equation}
Let us remark that on the right-hand side of the equations, there are
already $\beta$-expansions.

Repeating the rules (\ref{1}) and (\ref{2}), it is possible to
transform any finite $\beta$-representation of a~number $x$ into the
$\beta$-expansion of $x$. As we reduce the sum of digits in the
$\beta$-representation by applying rules (\ref{1}) and (\ref{2}),
after a~finite number of steps we get the $\beta$-expansion.

\begin{example}
$ \ (p+2)q(q+1)\bullet \ = \ (p+1)00\bullet \ + \ 1q(q+1)\bullet \ =
\ 10(p-q)0\bullet \ + \ 1q(q+1)\bullet \ = \ 11p(q+1)\bullet \ = \
1200\bullet (p-q)$
\end{example}
On the other hand, the rules (\ref{1}) and (\ref{2}) raise the sum
of digits on the right-hand side of the fractional point $\bullet$.
It means that the number of digits in the $\beta$-expansion of $x$
on the right-hand side of $\bullet$ is greater or equal to the
number of digits in any $\beta$-representation of $x$.

Therefore, the following fact holds true.
\begin{observation}\label{obs}
If $x,y \geq 0$ and $x,y \in Fin(\beta)$, then $fp_{\beta}(x+y)\geq
fp_{\beta}(x).$
\end{observation}

The following lemma is the most important tool to estimate
$L_{\oplus}(\beta)$.
\begin{lemma}\label{essential}
Let $x_kx_{k-1}\cdots x_0\bullet$ be the $\beta$-expansion of a~positive
$\beta$-integer $x$ and let $l \in \mathbb N$. Then
$x+{\beta}^l \in \mathbb Z_{\beta}$ or there exists $s \geq l$ such
that
\begin{enumerate}
\item for $l=0$,
$$\langle x+{\beta}^l \rangle_{\beta}=x_k\cdots(x_{s+1}+1)0^{s+1}\bullet (p-q),$$
\item for $l \geq 1$,
$$\langle x+{\beta}^l \rangle_{\beta}=x_k\cdots(x_{s+1}+1)0^{s-l+1}(x_{l-1}-q)
\cdots(x_1-q)(x_0-q-1)\bullet (p-q).$$
\end{enumerate}
\end{lemma}
\begin{proof}
\begin{enumerate}
\item For $l=0$.
Let us suppose that $x+{\beta}^0=x+1 \not \in \mathbb Z_{\beta}$.
Then $x_kx_{k-1}\cdots(x_0+1)\bullet$ is not a~$\beta$-expansion of
$x+1$. Therefore the suffix has the form $(p+1)$ or $pq^{s-1}(q+1),$
where $s \geq 1$. Applying the rule (\ref{1}), resp. (\ref{2}), the
$\beta$-representation of $x+1$ can be rewritten as
$$x_kx_{k-1}\cdots x_1(p+1)\bullet \ = \ x_k\cdots x_2(x_1+1)0\bullet (p-q)$$
or
$$x_kx_{k-1}\cdots x_{s+1}pq^{s-1}(q+1)\bullet \ = \ x_k\cdots(x_{s+1}+1)0^{s+1}\bullet (p-q).$$
Now, it suffices to show that the expressions on the right-hand side
are already $\beta$-expansions, or, equivalently, they fulfill the
Parry condition. It follows immediately from the fact that if
$x_k\cdots x_1p$ and $x_k\cdots x_{s+1}pq^s$ fulfill the Parry condition,
then $x_k\cdots(x_1+1)0$ and $x_k\cdots(x_{s+1}+1)0^{s+1}$ fulfill this
condition, too.
\item For $l \geq 1$. Let us suppose that $x+{\beta}^l \not \in
\mathbb Z_{\beta}$. Then
\begin{equation} \label{l1}
x_k\cdots x_{l+1}(x_l+1)x_{l-1}\cdots x_0
\end{equation}
does not fulfill
the Parry condition. There can be three reasons for that.
\begin{enumerate}
\item $x_l=q-1$,
\item $x_l=p$,
\item $x_l=q$.
\end{enumerate}
 \begin{enumerate}
   \item Let $x_l=q-1$. Denote $s=\min\{i>l \bigm | x_i=p\}$.
   Obviously, $x_i=q$ for all $i$, $s>i>l$. Necessarily, $x_{s+1}<p$.
   If we knew that for all $i<l$ it holds $x_i \geq q$ and $x_0\geq
   q+1$, then we could apply the rule (\ref{2}) for rearranging the
   $\beta$-representation of $x+{\beta}^l$ in the following way:
   $$x_k\cdots x_{s+1}pq^{s-l}x_{l-1}\cdots x_0\bullet \ =$$
   $$ (x_{l-1}-q)\cdots(x_1-q)(x_0-q-1)\bullet \ + \ x_k\cdots x_{s+1}pq^{s-1}(q+1)\bullet \ =$$
   $$(x_{l-1}-q)\cdots(x_1-q)(x_0-q-1)\bullet \ + \ x_k\cdots(x_{s+1}+1)0^{s+1}\bullet (p-q) \ =$$
   $$ x_k\cdots(x_{s+1}+1)0^{s-l+1}(x_{l-1}-q)\cdots(x_1-q)(x_0-q-1)\bullet (p-q)$$
   Since the last expression fulfills
   the Parry condition, we have obtained the $\beta$-expansion of
   $x+{\beta}^l$. Let us show that the conditions $x_0\geq q+1$ and
   $x_i \geq q$ for all $i<l$ are always true. Firstly, we prove that
   $x_i \geq q$ for all $i<l$. Let us prove it by contradiction. Let us
   denote by $i_0$ the maximal index $<l$ such that $x_{i_0}\leq q-1$.
   Then, let us denote by $j_0$ the minimal index $>i_0$ such that
   $x_{j_0}\geq q+1$. Such an index exists because (\ref{l1}) does not
   fulfill the Parry condition. Hence, the chain (\ref{l1}) has the
   following form:
   $$x_k\cdots x_{s+1}pq^{s-l}x_{l-1}\cdots x_{j_0+1}x_{j_0}q^{j_0-i_0-1}x_{i_0}x_{i_0-1}\cdots x_0$$
   Using the rule (\ref{2}), we get the $\beta$-representation of
   $x+{\beta}^l$ in the form:\\
   if $j_0>i_0+1$,
   $$x_k\cdots (x_{s+1}+1)0^{s-l+1}(x_{l-1}-q)\cdots(x_{j_0+1}-q)(x_{j_0}-q-1)pq^{j_0-i_0-2}x_{i_0}x_{i_0-1}\cdots x_0\bullet$$
   if $j_0=i_0+1$,
   $$x_k\cdots(x_{s+1}+1)0^{s-l+1}(x_{l-1}-q)\cdots(x_{j_0+1}-q)(x_{j_0}-q-1)(x_{i_0}+p-q)x_{i_0-1}\cdots x_0\bullet$$
   In both cases, these $\beta$-representations are already the
   $\beta$-expansions, thus we get a~contradiction with the fact that
   $x+{\beta}^l \not \in \mathbb Z_{\beta}$. Secondly, we show that
   $x_0\geq q+1$. Let us prove it again by contradiction. Let us
   suppose that $x_0=q$, then there exists $t\geq 1$ such that $q^t$
   is the suffix of the chain $x_k\cdots x_0$. Let us consider the maximal
   such $t$. Then the $\beta$-representation of $x+{\beta}^l$ has the
   following form:
   $$x_k\cdots x_{s+1}pq^{s-l}x_{l-1}\cdots x_{t+1}x_tq^t\bullet$$
   where $x_i \geq q$ for all $i \in \{t+1, \ldots, l-1\}$ and $x_t\geq
   q+1$. Applying the rule (\ref{2}), we can rewrite the
   $\beta$-representation as
   $$x_k\cdots(x_{s+1}+1)0^{s-l+1}(x_{l-1}-q)\cdots(x_{t+1}-q)(x_t-q-1)pq^{t-1}\bullet$$
   which is a~contradiction with $x+{\beta}^l \not \in \mathbb
   Z_{\beta}$.
\item Let $x_l=p$. Then $x_{l+1}<p$ and $x_{l-1}\leq q$. Using the
rule (\ref{1}), we obtain
\begin{equation}\label{p}
x_k\cdots x_{l+1}(p+1)x_{l-1}\cdots x_0\bullet \ = \
x_k\cdots(x_{l+1}+1)0(x_{l-1}+p-q)x_{l-2}\cdots x_0\bullet
\end{equation}
Since $x_l x_{l-1}\cdots x_0=px_{l-1}\cdots x_0 \preceq pq^{\omega}$, we
have $x_{l-1}\cdots x_0\preceq q^{\omega}$, and, consequently,
$(x_{l-1}+p-q)x_{l-2}\cdots x_0 \preceq pq^{\omega}$. Thus, the
expression on the right-hand side of (\ref{p}) is already the
$\beta$-expansion of $x+{\beta}^l$, which is a~contradiction with
$x+{\beta}^l \not \in \mathbb Z_{\beta}$.
\item Let $x_l=q$, then there exists $t\geq l$ such that
$x_k\cdots x_0=x_k\cdots x_{t+1}pq^{t-l}x_{l-1}\cdots x_0$. The
$\beta$-representation of $x+{\beta}^l$ equal to
$x_k\cdots x_{t+1}pq^{t-l-1}(q+1)x_{l-1}\cdots x_0\bullet$ can be rewritten,
using the rule (\ref{2}), as
$$x_k\cdots(x_{t+1}+1)0^{t-l+1}(x_{l-1}+p-q)x_{l-2}\cdots x_0\bullet$$
which is already the $\beta$-expansion of $x+{\beta}^l$. Thus, we
arrive again at a~contradiction with $x+{\beta}^l \not \in \mathbb
Z_{\beta}.$
\end{enumerate}
\end{enumerate}
\end{proof}

\begin{proposition}\label{q}
Let $x,y \in \mathbb Z_{\beta}$, $x \geq y \geq 0$, and let all
digits in the $\beta$-expansion of $y$ be $\leq q$. Then the
$\beta$-fractional part of $x+y$ is either $0$ or $\frac{p-q}{\beta}$.
\end{proposition}
\begin{proof}
We will proceed by induction on the positive elements of $\mathbb
Z_{\beta}$. For $y=1$, the statement follows from
Lemma~\ref{essential}, as well as for $y=2,\ldots,q$. Let $y\geq q+1$,
$\langle y \rangle_{\beta}=y_ly_{l-1}\cdots y_0\bullet,$ where $y_l\geq
1$ and $y_i \leq q$ for all $i \in \{0,\ldots,l\}$. If $x+{\beta}^l \in
\mathbb Z_{\beta}$, then $x+y=\tilde{x}+\tilde{y}$, where
$\tilde{x}=x+{\beta}^l$ and $\tilde{y}=y-{\beta}^l$, and the
statement follows by applying the induction assumption on
$\tilde{y}=y-\beta^l < y$. If $x+{\beta}^l \not \in \mathbb
Z_{\beta}$, then using Lemma~\ref{essential}, we get
\begin{equation}\label{3}
x+y=x+{\beta}^l+(y-{\beta}^l)=x_k\cdots(x_{s+1}+1)0^{s-l}(y_l-1)(x_{l-1}+y_{l-1}-q)\cdots(x_0+y_0-q-1)\bullet
(p-q)
\end{equation}
According to Lemma~\ref{essential},
$x_k\cdots(x_{s+1}+1)0^{s-l+1}(x_{l-1}-q)\cdots(x_0-q-1)\bullet (p-q)$ is
the $\beta$-expansion of $x+{\beta}^l$. Moreover, $y_l-1 \leq q-1$
and $(x_{l-1}+y_{l-1}-q)\cdots(x_0+y_0-q-1)\preceq x_{l-1}\cdots x_0$.
Consequently, the right-hand side of (\ref{3}) is already the
$\beta$-expansion of $x+y$.
\end{proof}
It is known that if $d_{\beta}(1)$ is eventually periodic, then the
set $Fin(\beta)$ is not closed under subtraction of positive
elements. In our case, we have for instance: $\beta-1=(p-1)\bullet
q^{\omega}$.
\begin{observation}\label{obs2}
Let $x \geq y \geq 0$, $x,y \in \mathbb Z_{\beta}$, then $x-y \in
\mathbb Z_{\beta}$ or $x-y \not \in Fin(\beta)$.
\end{observation}
\vspace{-0.2 cm} To prove this statement by contradiction one
assumes that $x-y \in Fin(\beta)-\mathbb Z_{\beta}$, i.e.
$fp_{\beta}(x-y)\geq 1$. Observation~\ref{obs} implies that
$fp_{\beta}(x)=fp_{\beta}(x-y+y)\geq fp_{\beta}(x-y)\geq 1$ and it
is a~contradiction with $x \in \mathbb Z_{\beta}$.

\begin{thm}
Let $d_{\beta}(1)=pq^{\omega}$. Then $L_{\oplus}(\beta)\leq
\lceil\frac{p}{q}\rceil.$
\end{thm}
\begin{proof}
Let $x,y \in \mathbb Z_{\beta}$ and $x,y \geq 0$. If $x-y \in
Fin(\beta)$, then necessarily $fp_{\beta}(x-y)=0$, as we have
mentioned in Observation~\ref{obs2}. Consequently, it suffices to
consider the addition $x+y$. Without loss of generality, we can
limit to the case $x \geq y$. Apparently, $y$ can be written as:
$$y=y^{(1)}+y^{(2)}+\cdots+y^{(s)},$$
where $s \leq \lceil\frac{p}{q}\rceil$ and the digits of $y^{(i)}$
are $\leq q$ for all $i=1,\ldots,s$. According to Proposition~\ref{q},
if we add to a~number of $Fin(\beta)$ a~$\beta$-integer with small
digits, the length of fractional part increases at most by 1. This
proves the statement.
\end{proof}
As an immediate consequence of the previous proof, we have the
following corollary.
\begin{corollary} \label{plus}
Let $x,y \in \mathbb Z_{\beta}$ and $x,y \geq 0$. Then there
exists $\varepsilon \in \{0,1,\ldots,\lceil\frac{p}{q}\rceil \}$
such that
$$x+y \in \mathbb Z_{\beta}+\varepsilon \,\frac{~p\!-\!q~}{\beta}\
.$$
\end{corollary}

\subsection{An upper bound on the constant $c$ in the balance property of
$u_{\beta}$}\label{implicationbalance}

Corollary~\ref{plus} allows us to derive an upper bound on the
balance function of $u_{\beta}$. Let us remind that $u_{\beta}$
arises if we associate with the longer gap between neighboring
$\beta$-integers the letter $A$ and with the shorter one the
letter $B$. The length of the longer gap is $\Delta_A=1$ and of
the shorter one $\Delta_B=1-\frac{p-q}{\beta}$.
\begin{proposition}\label{u_beta max A}
$u_{\beta}$ is $\lceil\frac{p}{q}\rceil $-balanced. Moreover, any prefix of $u_{\beta}$ contains at least the same number of letters $A$ as any other factor of $u_{\beta}$ of the same length.
\end{proposition}
\begin{proof}
Let $w$ be a~factor of $u_{\beta}$ of the length $n$ and $\hat w$ be the prefix of $u_{\beta}$ of the same length. Find $\beta$-integers $x$ and $y$, $x<y$, such that the sequence of distances between neighboring $\beta$-integers in the segment of $\mathbb Z_{\beta}$ from $x$ to $y$ corresponds to the factor $w$. Clearly,
\begin{equation}\label{A}
y=x+|w|_A\Delta_A+|w|_B\Delta_B.
\end{equation}
The prefix $\hat w$ corresponds to the $\beta$-integer
\begin{equation} \label{B}
z=|\hat w|_A\Delta_A+|\hat w|_B\Delta_B.
\end{equation}
Corollary~\ref{plus} implies that there exists $\hat z \in \mathbb Z_{\beta}$ such that
\begin{equation}\label{C}
x+z=\hat z+\varepsilon(\Delta_A-\Delta_B), \ \mbox{for }\ \varepsilon \in \{0,1,\ldots,\lceil\frac{p}{q}\rceil \}.
\end{equation}
Since $y,\hat z \in \mathbb Z_{\beta}$, it is possible to express the distance between $y$ and $\hat z$ as a~combination of the lengths of gaps $\Delta_A$ and $\Delta_B$, i.e.
there exist $L,M \in \mathbb N$ such that
\begin{equation} \label{D}
\hat z -y= \pm (L\Delta_A+M\Delta_B).
\end{equation}
Using (\ref{A}), (\ref{B}), and (\ref{C}), we get
$$\hat z-y=x+z-\varepsilon(\Delta_A-\Delta_B)-x-|w|_A\Delta_A-|w|_B\Delta_B=$$
$$(|\hat w|_A-|w|_A-\varepsilon)\Delta_A+(|\hat w|_B-|w|_B+\varepsilon)\Delta_B=$$
\begin{equation}\label{E}
(|\hat w|_A-|w|_A-\varepsilon)\Delta_A-(|\hat w|_A-|w|_A-\varepsilon)\Delta_B
\end{equation}
In the last equation, we have used the fact that the factors $w$ and
$\hat w$ have the same lengths, and, consequently, $|\hat
w|_A-|w|_A=|w|_B-|\hat w|_B$. As $\Delta_A=1$ and
$\Delta_B=1-\frac{p-q}{\beta}$ are linearly independent over
$\mathbb Q$, the expression of $\hat z-y$ in (\ref{E}) as an integer
combination of the lengths of gaps is unique. Since $L, M$ are
nonnegative, from (\ref{D}) and (\ref{E}) it follows that $|\hat
w|_A-|w|_A-\varepsilon=0$, i.e.
$$|\hat w|_A=|w|_A+\varepsilon,$$
where $\varepsilon \in  \{0,1,\ldots,\lceil\frac{p}{q}\rceil \},$
which is exactly the statement of the proposition.
\end{proof}

\section{Balance property of $u_{\beta}$}\label{balancebound}
In the previous section, we have proved, using arithmetical
properties of $\beta$-integers, that the infinite word $u_{\beta}$
is  $\lceil\frac{p}{q}\rceil$-balanced. In this section, we will
even show that $u_{\beta}$ is $\lceil\frac{p-1}{q}\rceil$-balanced,
which is a~better estimate in the case when $q$ divides $p-1$. We
will as well prove that this estimate cannot be improved. As
a~consequence, this result will be used to obtain a~lower bound on
$L_{\oplus}(\beta)\geq \lfloor\frac{p-1}{q}\rfloor.$


At first, we state without any proof some trivial properties of
the fixed point $u_\beta$ of the substitution. Let us recall it:
$$A\mapsto A^pB, \qquad
B\mapsto A^qB, \qquad \hbox{for} \ \  p>q>1\,.$$


\begin{observation}\label{o3}
Let $BA^kB$ be a factor
of $u_\beta$. Then $k=p$ or $k=q$. In particular, if $A^k$ is a factor of $u_\beta$, then $k\leq p$.
\end{observation}

\begin{observation}\label{exist w}
If $v$ is a finite factor of $u_\beta$, then $B\varphi(v)$ is also a factor of $u_\beta$.
\end{observation}

\begin{observation}\label{vzor}
Let $BvB$ be a factor of $u_\beta$. Then there exists a unique
factor $w$ of $u_\beta$ such that $vB=\varphi(w)$.
\end{observation}


Now we describe two sequences of  factors of $u_\beta$ denoted by
$\left (w_\beta^{(n)}\right)_{n=1}^\infty$ and $\left
(u_\beta^{(n)}\right)_{n=1}^\infty$, whose behaviour fully
determines the balance properties of $u_\beta$.

Let us define a~sequence $\left
(w_\beta^{(n)}\right)_{n=1}^\infty$ recursively by
\begin{equation}\label{defw}
\begin{array}{rcll}
w_\beta^{(1)}&=&B\\
w_\beta^{(n)}&=&B\varphi(w_\beta^{(n-1)}) &\hbox{ for } n\in\mathbb
N,\ n\geq2.
\end{array}
\end{equation}
According to Observation~\ref{exist w} the words $w_\beta^{(n)}$
are factors of $u_\beta$. Note that the sequence
$\left(|w_\beta^{(n)}|\right)_{n=1}^\infty$ is strictly
increasing.

Furthermore, we define sequence $\left
(u_\beta^{(n)}\right)_{n=1}^\infty$ by
  $$u_\beta^{(n)} = \hbox{prefix of \ } u_\beta \  \hbox{of the
  length}
  \ |w_\beta^{(n)}|\,.$$


\begin{observation}\label{w}
For all $n\in\N, n \geq 1$,
$$
w_\beta^{(n+1)}=w_\beta^{(n)}\hat{u}^{(n)}B,
$$
where $\hat{u}^{(n)}$ is a~prefix of $u_\beta$.
\end{observation}

\pf
By induction on $n$:

For $n=1$, we have $w_\beta^{(2)}=B\varphi(w_\beta^{(1)})=
B\varphi(B)=BA^qB=w_\beta^{(1)}A^qB$;\quad $\hat{u}^{(1)}=A^q$.

Suppose that $w_\beta^{(n)}=w_\beta^{(n-1)}\hat{u}^{(n-1)}B$ and
$\hat{u}^{(n-1)}$ is a~prefix of $u_\beta$. Then
$$
w_\beta^{(n+1)}=B\varphi(w_\beta^{(n)})=B\varphi(w_\beta^{(n-1)}\hat{u}^{(n-1)}B)=
B\varphi(w_\beta^{(n-1)})\varphi(\hat{u}^{(n-1)})A^qB=w_\beta^{(n)}\hat{u}^{(n)}B\,,
$$
where $\hat{u}^{(n)}=\varphi(\hat{u}^{(n-1)})A^q$ is a~prefix of
$u_\beta$ according to Observation~\ref{o3}.

\pfk

Observation~\ref{w} allows us to define an infinite word $w_\beta$ in ${\mathcal A}$ as
$$
w_\beta=\lim_{n\to\infty}w_\beta^{(n)}.
$$
It follows from the definition of  $w_\beta^{(n)}$ that this
infinite word fulfils
\begin{equation}\label{nekonecne}
w_{\beta}=B\varphi(w_{\beta}).
\end{equation}
Consequently, using
Observation~\ref{vzor} we get the following observation.

\begin{observation}\label{vzorw}
Let $w'B$ be a prefix of $w_\beta$. Then the unique factor $w''$
of $u_\beta$ satisfying $w'B=B\varphi(w'')$  is a prefix of
$w_\beta$.
\end{observation}

We know already from Proposition~\ref{u_beta max A} that prefixes of
$u_{\beta}$ are factors
with the largest number of letters $A$.
The infinite word $w_\beta$ plays the same role for letters $B$.

\begin{proposition}\label{w max B}
Any prefix of $w_{\beta}$ contains at least the same number of
letters $B$ as any other factor of the same length.
\end{proposition}

\pf We will prove the statement by contradiction. Let us assume that
there exist a~$k\in\N$ and a~factor $v=v_0v_1v_2\cdots v_{k-1}$ of
$u_\beta$ such that $|w|_B<|v|_B$, where $w=w_0w_1w_2\cdots w_{k-1}$
is a prefix of $w_\beta$. We choose the minimal $k$ with this
property. Then
\begin{equation}\label{w:balance}
|v|_B=|w|_B+1.
\end{equation}
Minimality of $k$ implies that $v_0=B$, $v_{k-1}=B$ and
$w_{k-1}=A$. The fact that $w$ is a prefix of $w_\beta$ which
satisfies \eqref{nekonecne}, implies $w_0 = B$. Thus
$v_{k-1-q}v_{k-q}\cdots v_{k-3}v_{k-2}=A^q$ according to
Observation~\ref{o3}, hence $w_{k-1-q}w_{k-q}\cdots
w_{k-3}w_{k-2}=A^q$ by virtue of minimality of $k$.
Observation~\ref{o3} together with the fact $w_{k-1}=A$ imply that
there is a uniquely determined integer $j$ satisfying $0\leq j\leq
p-q-1$ such that $wA^jB$ is a factor of $u_\beta$. Since $v_0=B$,
$w_0=B$ and $v_{k-1}=B$, we may use Observation~\ref{vzor} to
deduce that there are unique factors $v'$ and $w'$ of $u_\beta$
such that $\varphi(v')=v_1v_2\cdots v_{k-1}$ and
$\varphi(w')=w_1w_2\cdots w_{k-1}A^jB, k\geq 1$. Since
$\varphi(v')$ and $\varphi(w')$ contain the same number of letters
$B$, clearly $|v'| = |w'| < k$.
Moreover, it follows from
Observation~\ref{vzorw} that the factor $w'$ is a~prefix of
$w_\beta$. As $\varphi(v')$ is shorter than $\varphi(w')$, the
word $v'$ contains more letters $B$ than $w'$, which is a~prefix
of $w_\beta$. It is a~contradiction with the minimality of $k$.
\pfk

\begin{lemma}\label{zjednoduseni}
Let $v$, $v'$ be factors of $u_\beta$ of the same length $k$, let
$n$ be such a~positive integer that $|w_\beta^{(n)}|\leq
k<|w_\beta^{(n+1)}|$. Then
$$
\left| |v|_B-|v'|_B \right|\leq |w_\beta^{(n)}|_B-|u_\beta^{(n)}|_B\,.
$$
\end{lemma}

\pf
Propositions~\ref{u_beta max A} and~\ref{w max B} imply
$$
\left| |v|_B-|v'|_B \right| \leq |w'|_B-|u'|_B\,,
$$
where $u'$ and $w'$ are prefixes of $u_\beta$ and $w_\beta$, respectively, of length $k$.
Observation~\ref{w} together with the assumption $k<|w_\beta^{(n+1)}|$ implies that $w'=w_\beta^{(n)}\hat{u}$ for some prefix $\hat{u}$ of $u_\beta$. Let us write the factor $u'$ in the form $u'=u_\beta^{(n)}\hat{v}$. Using Proposition~\ref{u_beta max A}, we get
$$
|w'|_B-|u'|_B=|w_{\beta}^{(n)}|_B-|u_{\beta}^{(n)}|_B+|\hat{u}|_B-|\hat{v}|_B\leq
|w_\beta^{(n)}|_B-|u_\beta^{(n)}|_B\,,
$$
which concludes the proof of the statement.
\pfk

Lemma~\ref{zjednoduseni} will be very useful in the investigation of
balance properties, since it enables us to find out the optimal
balance bound of the word $u$ from examining the sequence
$(D_n)_{n=1}^{\infty}$, where
$$D_n:=|w_\beta^{(n)}|_B-|u_\beta^{(n)}|_B.$$



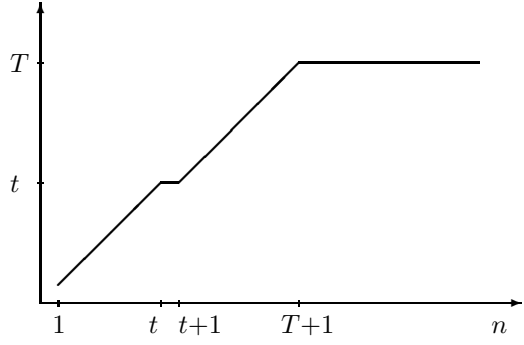
\begin{figure}[!h]
\begin{center}
\setlength{\unitlength}{0.8cm}
\begin{picture}(12,8)
\put(1,3){\vector(1,0){8}} \put(1,3){\vector(0,1){5}}
\put(8.5,2.5){\mbox{$n$}} \put(3,2.95){\line(0,1){0.1}}
\put(1.3,2.95){\line(0,1){0.1}} \put(0.95,5){\line(1,0){0.1}}
\put(2.8,2.5){\mbox{$t$}} \put(1.2,2.5){\mbox{1}}
\put(0.5,4.85){\mbox{$t$}}
\thicklines \put(1.3,3.3){\line(1,1){1.7}} \thinlines
\put(3.3,2.95){\line(0,1){0.1}} \put(3.3,2.5){\mbox{$t$+1}}
\thicklines \put(3,5){\line(1,0){0.3}} \thinlines
\put(5.3,2.95){\line(0,1){0.1}} \put(0.95,7){\line(1,0){0.1}}
\put(5,2.5){\mbox{$T$+1}} \put(0.5,6.85){\mbox{$T$}}
\thicklines \put(3.3,5){\line(1,1){2}} \put(5.3,7){\line(1,0){3}}
\end{picture}
\vskip-1cm
\caption{$t=\lfloor \frac{p+q}{q+1} \rfloor$ and $T=\lceil \frac{p-1}{q}\rceil$.}\label{sequenceDn}
\end{center}
\end{figure}

In the sequel, we will show that the sequence $(D_n)$ has the form
depicted in Figure ~\ref{sequenceDn}, which shows that $u_\beta$ is
$\lceil \frac{p-1}{q}\rceil$-balanced and that this bound cannot be
diminished.

 To determine the value of $D_{n+1}$ using the value
of $D_n=|w_{\beta}^{(n)}|_B- |u_{\beta}^{(n)}|_B$, it is important
to take in account that:
\begin{enumerate}

\item Since the number of letters $A$ in the word
$u_{\beta}^{(n)}$ is by $D_n$ greater than in $w_{\beta}^{(n)}$,
the length of $\varphi(u_{\beta}^{(n)})$ is by $(p-q)D_n$ letters
longer than the length of $\varphi(w_{\beta}^{(n)})$.

\item $w_{\beta}^{(n+1)}=B\varphi(w_{\beta}^{(n)})$.

\item $u_{\beta}^{(n+1)}$ is a~prefix of $u_\beta$ chosen so that
$|u_{\beta}^{(n+1)}|=|w_{\beta}^{(n+1)}|$.

\item Since $u_\beta$ is the fixed point of the substitution,
$\varphi(u_{\beta}^{(n)})$ is a~prefix of $u_\beta$ as well.

\item $u_{\beta}^{(n+1)}$ can be obtained from
$\varphi(u_{\beta}^{(n)})$ by erasing of its suffix of length
$(p-q)D_n-1$.

\item As the lengths of $w_{\beta}^{(n)}$ and $u_{\beta}^{(n)}$
are the same, $\varphi(w_{\beta}^{(n)})$ and
$\varphi(u_{\beta}^{(n)})$ contain the same number of letters $B$.

\end{enumerate}

\noindent These six  simple facts imply the following recurrence
relation for the sequence $(D_n)$:

\begin{equation}\label{Dn}
D_{n+1}=1+ |v|_B \,,\ \  \hbox{where} \  v \ \hbox{is a suffix of
}   \  \varphi(u_\beta^{(n)}) \ \hbox{ and  }  \ |v| = (p-q)D_n-1
\end{equation}

\noindent Consequently, to determine the value of $D_{n+1}$, one
needs to know the form of the suffix of
$\varphi(u_{\beta}^{(n)})$, hence the form of the suffix of
$u_{\beta}^{(n)}$.

\begin{proposition}\label{formDn}
Let us denote by $t:=\lfloor \frac{p+q}{q+1}\rfloor$ and $T=\lceil
\frac{p-1}{q}\rceil$.
\begin{enumerate}
\item
If $n \leq t$, then $D_n=n$ and $u_{\beta}^{(n)}$ has the suffix $A^{(n-1)q+n}$.
\item
If $t+1\leq n \leq T+1$, then $D_n=n-1$ and the suffix of
$u_{\beta}^{(n)}$ is $A^pBA^{(n-1)(q+1)-p}$.
\item
If $T+1\leq n$, then $D_n=T$ and $u_{\beta}^{(n)}$ has the suffix $A^{T-1}$.

\end{enumerate}
\end{proposition}
\begin{proof}
Let us show how the statement $3.$ follows from $2.$
For $n=T+1$ the statement $2.$ implies that
\begin{equation} \label{0inequality}
u_{\beta}^{(n)} \ \mbox{has the suffix} \ A^{T-1} \ \mbox{and}  \
D_n=T.
\end{equation}
Let us use the rule~(\ref{Dn}) to calculate $D_{n+1}$. The word
$\varphi(u_{\beta}^{(n)})$ has the suffix $$\varphi(A^{T-1})=
\underbrace{(A^pB)(A^pB)\dots (A^pB)}_{(T-1)-\mbox{times}}.$$ We
erase from this word the suffix of length $(p-q)T-1$. Let us show
that in this procedure we have erased all the letters $B$,
i.e. $T-1$ letters $B$. To verify this statement, it suffices to
prove the inequality
\begin{equation}\label{1inequality}
(p-q)T-1 \geq (p+1)(T-2)+1.
\end{equation}
In order to prove that by erasing of the suffix of length $(p-q)T-1$, there are still at least
$T-1$ letters left in the word $\varphi(A^{T-1})$, one has to show
\begin{equation}\label{2inequality}
(p+1)(T-1)-(p-q)T+1 \geq T-1.
\end{equation}
Consequently, if we verify the equations~(\ref{1inequality})
and~(\ref{2inequality}), it will be proved that $D_{n+1}=T$ and
$u_{\beta}^{(n+1)}$ has the suffix $A^{T-1}$. It means by virtue
of~(\ref{0inequality}) for index $n$, we have shown the virtue for
index $n+1$, thus, using induction, for all $n \geq T+1$.

\medskip

Equation~(\ref{2inequality}) is equivalent to $T \geq
\frac{p-1}{q}$, which is evidently true for the choice of
$T=\lceil \frac{p-1}{q}\rceil$. Equation~(\ref{1inequality}) is
equivalent to $T \leq \frac{2p}{q+1}$. By means of the fact which
holds for positive integers $a,b$ $$\left\lceil
\frac{a}{b}\right\rceil \leq \frac{a}{b}+\frac{b-1}{b}\,,$$ we get
$$T=\left \lceil \frac{p-1}{q}\right \rceil \leq \frac{p-1}{q}+\frac{q-1}{q}=\frac{p+q-2}{q}\,.$$
It is enough to verify that $\frac{p+q-2}{q} \leq \frac{2p}{q+1}$,
which is equivalent with $(q+1)(q-2) \leq p(q-1).$ This equation holds because in our substitution
$p \geq q+1$.

The validity of $1.$ and $2.$ can be shown analogically by induction on $n$.
\end{proof}

As an immediate consequence of the recently proven proposition,
we have the following essential theorem.
\begin{thm}
The infinite word $u_\beta$ is $c$-balanced, where $c=\lceil \frac{p-1}{q} \rceil$.
This value $c$ is the smallest possible.
\end{thm}

\subsection{A lower bound on $L_{\oplus}(\beta)$}
To derive a~lower bound on $L_{\oplus}(\beta)$, we will use the
fact that there exist a~factor $w$ and a~prefix $\hat w$ of
$u_{\beta}$ such that $|\hat
w|_A=|w|_A+\lceil\frac{p-1}{q}\rceil.$ Let $x,y \in \mathbb
Z_{\beta}$, $x<y$, such that the gaps in the segment of $\mathbb
Z_{\beta}$ from $x$ to $y$ correspond to the word $w$. And, let $z
\in \mathbb Z_{\beta}$ be the $\beta$-integer corresponding to the
prefix $\hat w$. Then
$$x+z=y+\Bigl\lceil\frac{p-1}{q}\Bigr\rceil(\Delta_A-\Delta_B)=y+\Bigl\lceil\frac{p-1}{q}\Bigr\rceil\frac{p-q}{\beta}.$$
From Observation~\ref{obs}, it follows that
$$fp_{\beta}(x+z)=fp_{\beta}\left(y+\Bigl\lceil\frac{p-1}{q}\Bigr\rceil\frac{p-q}{\beta}\right)\geq
fp_{\beta}\left
(\Bigl\lceil\frac{p-1}{q}\Bigr\rceil\frac{p-q}{\beta}\right)\geq
fp_{\beta}\left(\Bigl\lfloor\frac{p-1}{q}\Bigr\rfloor\frac{p-q}{\beta}\right).$$
Now, it suffices to show that $
fp_{\beta}\bigl(\lfloor\frac{p-1}{q}\rfloor\frac{p-q}{\beta}\bigr)=\lfloor\frac{p-1}{q}\rfloor.$

\begin{lemma} \label{F}
For $j=1,\ldots,\lfloor\frac{p-1}{q}\rfloor$, the
$\beta$-expansion of the number $j \frac{p-q}{\beta}$ is
$$\left\langle j \frac{p-q}{\beta}\right\rangle_{\beta}=(j-1)\bullet a_j\cdots a_1\,,$$
where $a_1:=p-q$ and $a_i:=(p-1)-iq$ for
$i=2,\ldots,\lfloor\frac{p-1}{q}\rfloor$.
\end{lemma}

\begin{proof}
The numbers $a_i$ are defined so that $a_i \geq 0$ and $(j-1)a_j a_{j-1}\cdots a_1 \prec pq^{\omega}$.
Thus, the expression $(j-1)\bullet a_j\cdots a_1$ is the $\beta$-expansion of a~positive number. Now,
we have to show that
$$j \frac{p-q}{\beta}=j-1+\frac{a_j}{\beta}+\frac{a_{j-1}}{{\beta}^2}+\cdots+\frac{a_1}{{\beta}^j},$$
which can be done easily by mathematical induction on $j$.
\end{proof}
Lemma~\ref{F} confirms that $
fp_{\beta}\bigl(\lfloor\frac{p-1}{q}\rfloor\frac{p-q}{\beta}\bigr)=\lfloor\frac{p-1}{q}\rfloor,
$ in other words, it implies the announced lower bound on
$L_{\oplus}(\beta)$. To sum up, we have derived the following
theorem.
\begin{thm}
Let $d_{\beta}(1)=pq^{\omega}$. Then
$$\left\lfloor\frac{p-1}{q}\right\rfloor  \leq L_{\oplus}(\beta) \leq
\left\lceil\frac{p}{q}\right\rceil.$$
\end{thm}

Let us mention that   difference between the upper bound
 $\lceil\frac{p}{q}\rceil$ and the lower bound
 $\lfloor\frac{p-1}{q}\rfloor$  is always 1.  Our computer
 experiments support the conjecture
 $L_{\oplus}(\beta)=\lfloor\frac{p-1}{q}\rfloor$.

\section*{Acknowledgements}

The authors acknowledge financial support by Czech Science
Foundation GA \v{C}R 201/05/0169, by the grant LC06002 of the
Ministry of Education, Youth, and Sports of the Czech Republic.

\end{document}